\PassOptionsToPackage{svgnames}{xcolor}
\documentclass[12pt]{article}

\usepackage{needspace,amsmath,amssymb,amsfonts,amsthm,setspace,tikz,dsfont,wrapfig,lipsum,natbib, array,colortbl,mathrsfs,pifont,wasysym,nicefrac,multicol,booktabs,makecell,multirow,thm-restate,caption, subcaption,setspace}
\usetikzlibrary{patterns}
\usepackage{etoolbox}
\usepackage{thmtools,thm-restate}
\usepackage{caption} 
\usepackage[all]{nowidow}
\usepackage[makeroom]{cancel}

\usepackage[letterpaper]{geometry}
\usepackage{hyperref}

\hypersetup{                                             
    pdffitwindow=true,                                  
    pdfstartview={XYZ null null 1.00},                   
    pdfnewwindow=true,                                   
    colorlinks=true,                                     
    linkcolor=DarkBlue,                                     
    citecolor=DarkBlue,                                     
    urlcolor=DarkBlue,                                      
    pdfauthor = {Gabriel Ziegler},
    pdfkeywords = {Informational Robustness, Common Belief in Rationality},
    pdftitle = {IRCBR},
    pdfsubject = {IRCBR},
    pdfpagemode = UseNone}

  \usepackage{pgfplots}
  \pgfplotsset{compat=newest}
  \usetikzlibrary{plotmarks}
  \usetikzlibrary{arrows.meta}
  \usepgfplotslibrary{patchplots}
  \usepackage{grffile}
  \usetikzlibrary{positioning}

\pgfplotsset{plot coordinates/math parser=false}
\newlength\figureheight
\newlength\figurewidth
\usepgfplotslibrary{colormaps}
\usetikzlibrary{pgfplots.colormaps} 

\usepackage{fancyhdr,ifthen}
\fancypagestyle{appendix}{
\fancyhead{}
\fancyhead[R]{\ifthenelse{\isodd{\value{page}}}{\thepage}{}}
\fancyhead[L]{\ifthenelse{\isodd{\value{page}}}{\textsc{}}{\thepage}}

\fancyfoot{}
}

\newtheorem{definition}{Definition}

\newtheorem{proposition}{Proposition}

\newtheorem{corollary}{Corollary}
\newtheorem{remark}{Remark}

\theoremstyle{definition}

\usepackage{titlesec}

\titleformat*{\section}{\large\scshape\centering}
\titleformat*{\subsection}{\scshape\centering}
\titleformat*{\subsubsection}{\itshape}
\titleformat*{\paragraph}{\large\bfseries\centering}
\titleformat*{\subparagraph}{\large\bfseries\centering}

\DeclareMathOperator*{\supp}{supp}

\DeclareMathOperator*{\marg}{marg}
\DeclareMathOperator*{\graph}{graph}
\DeclareMathOperator*{\argmax}{arg\,max}

\makeatletter
\DeclareRobustCommand\citepos
  {\begingroup
   \let\NAT@nmfmt\NAT@posfmt
   \NAT@swafalse\let\NAT@ctype\z@\NAT@partrue
   \@ifstar{\NAT@fulltrue\NAT@citetp}{\NAT@fullfalse\NAT@citetp}}

\let\NAT@orig@nmfmt\NAT@nmfmt
\def\NAT@posfmt#1{\NAT@orig@nmfmt{#1's}}

\makeatother

\title{\vspace{-1em}Informational Robustness of Common Belief in Rationality}

\author{
Gabriel Ziegler\footnote{The University of Edinburgh, School of Economics; 31 Buccleuch Place, Edinburgh, EH8 9JT, UK; \href{mailto:ziegler@ed.ac.uk}{\tt ziegler@ed.ac.uk}.
\newline
Enrico de Magistris, Marciano Siniscalchi, Peio Zuazo-Garin, and two anonymous referees provided very helpful comments, which I am thankful for.}
} 

\begin{document}

\maketitle
\thispagestyle{empty}
\vspace*{-2em}
\begin{abstract}
\small
In this note, I explore the implications of informational robustness under the assumption of common belief in rationality. That is, predictions for incomplete-information games which are valid across all possible information structures. First, I address this question from a global perspective and then generalize the analysis to allow for localized informational robustness.  

\vspace{0.2cm}

\end{abstract}
\cleardoublepage
\setcounter{page}{1}
\setstretch{1.3}
\section{Introduction}
\label{section:introduction}

For Bayesian games \cite{DFM07} introduce the solution concept of \emph{interim-correlated rationalizability (ICR)} and show that it captures the behavioral implications of common belief in rationality within the Bayesian game.\footnote{To be precise, it should be called \emph{rationality and} common \emph{correct} belief in rationality, but I will use the shorter wording throughout.} A crucial ingredient of a Bayesian game is the Harsanyi type space, which models player's information about the exogenous uncertainty and in addition implicitly determines the players' hierarchy of beliefs about this uncertainty. Changing the Harsanyi type space will change the predictions of ICR usually and therefore this solution concept is not informationally robust. In this note, I explore the robust behavioral implications when the analyst does not want to make the strong assumption about the exact information (and higher-order beliefs) players have about the exogenous uncertainty.

Early pioneers in the area of informational robustness include \cite{aumann87}, \cite{BD87}, and \cite{forges93,forges06}. \cite{BM13, BM16} recently exploited the full power of informational robustness to provide robust predictions in economic environments with uncertainty.\footnote{Some of these ideas are fruitfully applied to the theory of robust mechanism design as initiated by \cite{BM05,BM09,BM11}. Other papers dealing with related ideas about robustness include \cite{battigalli99, battigalli03}, \cite{BS03}, \cite{DFM07}, \cite{liu15}, \cite{tang15}, and \cite{GZ17}.} The main analysis here is closest to that of \cite{BM17}. However, there is major conceptual difference: their starting point is that play in a Bayesian game is governed by Bayes Nash equilibrium, while here I start from the primitive assumptions of common belief in rationality. I will discuss a natural connection between these two approaches along the way of the analysis.


\section{Main Analysis}
\label{section:3_main}

In this section, I will first introduce the relevant notation and definitions in \autoref{subsection:3_environment}. \autoref{subsection:3_inforobust_outside} establishes the main robustness result. A more general informational robustness question is addressed in \autoref{subsection:3_inforobust_delta}.


\subsection{The Economic Environment, Information Structures, and Solution Concepts}
\label{subsection:3_environment}

An \emph{economic environment} consists of a tuple $\mathcal{E}=\langle I,\Theta_0,(A_{i},\Theta_i,u_{i})_{i\in I}\rangle$ where $I=\{1,2\}$ is  set of \emph{players}\footnote{To simplify notation, I will consider two players throughout. All results can be extended to any finite number of players.} and $\Theta_0$ is a finite set of \emph{states of nature}. Moreover, for each player $i$, there is ($i$) a finite set of (pure) \emph{actions} $A_{i}$, ($ii$) a finite set of \emph{payoff types},\footnote{\cite{Magistris21} extends \autoref{proposition:union_ICR_BFR} to allow for more general action and type spaces.} and ($iii$) a \emph{utility function} $u_{i}:A \times \Theta\rightarrow\mathds{R}$, where $A:=\prod_{i\in I}A_{i}$ and $\Theta:=\prod_{i\in I \cup \{0\}}\Theta_{i}$  denote the set of action \emph{profiles} and \emph{payoff states}, respectively. For each player $i$, a randomization of own strategies $\alpha_{i}\in\Delta(A_{i})$ is referred to as a \emph{mixed} strategy,\footnote{Throughout this section, for any topological space $X$, as usual, $\Delta\left(X\right)$ denotes the set of probability measures on the Borel $\sigma$-algebra of $X$.} and a probability measure $\mu_{i}\in\Delta\left(\Theta_0 \times \Theta_{-i} \times A_{-i}\right)$, where $\Theta_{-i} := \Theta_{3-i}$ and $A_{-i}:=A_{3-i}$,\footnote{Similar notation will be used for indexing throughout.} as a \emph{conjecture}. When necessary, with some abuse of notation, $a_{i}$ refers to the degenerate mixed strategy that assigns probability one to pure action $a_{i}$. For each payoff type $\theta_i$, each conjecture $\mu_{i}$ and possibly mixed action $\alpha_{i}$ naturally induce \emph{expected} utility $U_{i}(\mu_{i};\alpha_{i}, \theta_i)$ and based on this, each player $i$'s \emph{best-reply} correspondence for payoff type $\theta_i$ is defined by assigning to each conjecture $\mu_{i}$ the subset of pure actions $BR_{i}(\mu_{i}; \theta_i)$ that maximize its corresponding expected utility.\footnote{That is, given conjecture $\mu_{i}$ and payoff type $\theta_i$ the expected utility is
\begin{align*} U_{i}(\mu_{i};\alpha_{i},\theta_i):=\sum_{(a_{i},a_{-i})\in A}\sum_{(\theta_0, \theta_{-i})\in \Theta_0 \times \Theta_{-i}}\mu_{i}[\theta_0, \theta_{-i}, a_{-i}]\cdot\alpha_{i}[a_{i}]\cdot u_{i}(a_{i},a_{-i},\theta_0,\theta_i, \theta_{-i})
\end{align*}
 for each possibly mixed action $\alpha_{i}$, and the set of best-replies is $BR_{i}(\mu_{i}; \theta_i):=\argmax_{a_{i}\in A_{i}}U_{i}(\mu_{i};a_i,\theta_i)$.} For the rest of this section, I consider the economic environment $\mathcal{E}$ to be fixed and therefore drop most explicit mentions to it.

\emph{Belief-free rationalizability (BFR)} is a solution concept due to \cite{battigalli99,battigalli03} that requires only the specification of an economic environment.\footnote{For a more recent discussion of the relationship of BFR to other solution concepts see \cite{BDGP11}. \cite{BM17} call this solution concept ex-post rationalizability because it is equivalent to the iterative deletion of ex-post dominated actions. \citeauthor{BM17} also define a slightly stronger solution concept which they call belief-free rationalizability. In their version there is a belief restriction in place constraining the supports of the beliefs. After \autoref{corollary:union_BNE}, I discuss their solution concept in more detail.} Action $a_{i}$ is \emph{belief-free rationalizable} for payoff type $\theta_i$ if it is iteratively a best-reply to a belief that assigns positive probability only to actions of the other player that survived the previous round. Formally, action $a_{i}$ is belief-free rationalizable for payoff type $\theta_i$ if $a_{i}\in BFR_{i}^{\infty}(\theta_i):=\bigcap_{n\geq 0}BFR_{i}^{n}(\theta_i)$, where $BFR_{i}^{0}(\theta_i):=A_{i}$ and inductively for any $n\in\mathds{N}$, 
\begin{align}\tag{$BFR^n$}\resizebox{0.8\textwidth}{!}{$
BFR_i^{n}(\theta_i):=\left\{a_{i}\in BFR_i^{n-1}(\theta_i)\left|
\begin{tabular}{r l}
\multicolumn{2}{l}{$\text{There exists  }\mu_{i}\in\Delta(\Theta_0 \times \Theta_{-i} \times A_{-i})\text{ s.t.:}$}\\[1ex]
$(i)$&$\supp \mu_{i}\subseteq \Theta_0 \times \graph\left( BFR_{-i}^{n-1}\right)$,\\[1ex]
$(ii)$&$a_{i}\in BR_{i}(\mu_{i}; \theta_i)$
\end{tabular}
\right.\right\}.$} \label{eq:BFR_iterate} 
\end{align}

\cite{BS99,BS02} show that BFR corresponds to the behavioral implications of common belief in rationality.\footnote{Their result is actually more general; see \cite{BDGP11}.} Usual arguments can be used to give a fixed-point definition of rationalizability:\footnote{\citet[Proposition 8.1]{battigalli03} establishes this equivalence in a more general setting.} for every player $i$ and every payoff type $\theta_i$ consider a set of actions $F_i(\theta_i)$ with the following fixed-point property
\begin{align}\tag{$BFR_{FP}$}
F_i(\theta_i):=\left\{a_{i}\in A_i\left|
\begin{tabular}{r l}
\multicolumn{2}{l}{$\text{There exists }\mu_{i}\in\Delta(\Theta_0 \times \Theta_{-i} \times A_{-i})\text{ s.t.:}$}\\[1ex]
$(i)$&$\supp \mu_{i}\subseteq \Theta_0 \times \graph\left(F_{-i}\right)$,\\[1ex]
$(ii)$&$a_{i}\in BR_{i}(\mu_{i}; \theta_i)$
\end{tabular}
\right.\right\}. \label{eq:BFR_FP}
\end{align}
Then the pair $\left(BFR_i^\infty\right)_{i \in I}$ understood as correspondences is equal to the pair of correspondences $\left(F_i\right)_{i \in I}$ satisfying the fixed-point property and are largest by set inclusion.

\citeauthor{harsanyi67}'s (\citeyear{harsanyi67}) approach differs from the previous approach by appending a type structure to the economic environment. For the purpose of this section, I will call such a type structure an \emph{information structure}.\footnote{Formally, there is no difference, but the interpretation changes. In this section, I want to model hard information, whereas \citeauthor{harsanyi67}'s type structures are only a modeling device to represent \emph{hierarchies of beliefs} about $(\theta_0, \theta_1, \theta_2)$. \citet[Section 1.1]{DS15} discuss the conceptual difference.\label{footnote:harsanyi_type_space}} An information structure is a tuple $\mathcal{Y}=\langle (Y_i, \pi_i)_{i\in I}\rangle$, where for each player $i$, $Y_i$ is a finite set of \emph{signal realizations} and thus $X_i:= \Theta_i \times Y_i$ is the set of \emph{information types} describing the private information of player $i$. Moreover, $\pi_i: X_i \rightarrow \Delta(\Theta_0 \times X_{-i})$ specifies the beliefs about the opponent's information and state of nature.\footnote{This definition of information structure is stated from an interim perspective. Up to measure-zero events, there is an equivalent formalization from an ex-ante perspective.}  An economic environment $\mathcal{E}$ together with an information structure $\mathcal{Y}$ constitutes a \emph{Bayesian game} $\mathcal{B}=\langle \mathcal{E}, \mathcal{Y}\rangle$.

\cite{DFM07} show that common belief in rationality within a Bayesian game is behaviorally characterized by \emph{interim correlated rationalizability}. Action $a_{i}$ is interim correlated rationalizable for information type $x_i = (\theta_i,y_i)$ if it survives an iterated elimination procedure similar to above, but where the beliefs about opponent's information and state of nature have to coincide with the information prescribed by $\pi_i(\theta_i,y_i)$. Formally, action $a_{i}$ is interim correlated rationalizable for information type $x_i = (\theta_i,y_i)$  if $a_{i}\in ICR_{i}^{\infty,\mathcal{Y}}(\theta_i,y_i):=\bigcap_{n\geq 0}ICR_{i}^{n,\mathcal{Y}}(\theta_i,y_i)$, where $ICR_{i}^{0,\mathcal{Y}}(\theta_i,y_i):=A_{i}$ and inductively for any $n\in\mathds{N}$,  
\begin{align}\tag{$ICR^n$}\resizebox{0.8\textwidth}{!}{$
ICR_i^{n,\mathcal{Y}}(\theta_i,y_i):=\left\{a_{i}\in ICR_i^{n-1,\mathcal{Y}}(\theta_i,y_i)\left|
\begin{tabular}{r l}
\multicolumn{2}{l}{$\text{There exists }\mu_{i}\in\Delta(\Theta_0 \times \Theta_{-i} \times Y_{-i} \times A_{-i})\text{ s.t.:}$}\\[1ex]
$(i)$&$\supp \mu_{i}\subseteq \Theta_0 \times \graph\left( ICR_{-i}^{n-1,\mathcal{Y}}\right)$,\\[1ex]
$(ii)$&$a_{i}\in BR_{i}(\marg_{\Theta_0 \times \Theta_{-i} \times A_{-i}}\mu_{i}; \theta_i)$,\\[1ex]
$(iii)$&$\marg_{\Theta_0 \times \Theta_{-i} \times Y_{-i}}\mu_{i} = \pi_i(\theta_i,y_i)$
\end{tabular}
\right.\right\}.$} \label{eq:ICR_iterate}
\end{align}
Again, this can be equivalently stated as fixed-point. \citet[Claim 3]{DFM07} establish this characterization:\footnote{Their setting slightly differs from the setting here, but their argument can easily be adapted.} for every player $i$ and every information type $(\theta_i, y_i)$ consider a set of actions $F_i(\theta_i,y_i)$ with the following fixed-point property
\begin{align}\tag{$ICR_{FP}$}\resizebox{0.8\textwidth}{!}{$
F_i(\theta_i,y_i):=\left\{a_{i}\in A_i\left|
\begin{tabular}{r l}
\multicolumn{2}{l}{$\text{There exists }\mu_{i}\in\Delta(\Theta_0 \times \Theta_{-i} \times Y_{-i} \times A_{-i})\text{ s.t.:}$}\\[1ex]
$(i)$&$\supp \mu_{i}\subseteq \Theta_0 \times \graph\left( F_{-i} \right)$,\\[1ex]
$(ii)$&$a_{i}\in BR_{i}(\marg_{\Theta_0 \times \Theta_{-i} \times A_{-i}}\mu_{i}; \theta_i)$,\\[1ex]
$(iii)$&$\marg_{\Theta_0 \times \Theta_{-i} \times Y_{-i}}\mu_{i} = \pi_i(\theta_i,y_i)$
\end{tabular}
\right.\right\}.$} \label{eq:ICR_FP}
\end{align}
Then the pair $\left(ICR^{\infty,\mathcal{Y}}_i\right)_{i \in I}$ understood as correspondences is equal to the pair of correspondences $\left(F_i\right)_{i \in I}$ satisfying the fixed-point property and are largest by set inclusion.

\begin{remark}\label{remark:infoonly_ICR}
	The notation might suggest that interim correlated rationalizability depends on the information structure $\mathcal{Y}$, but not on the economic environment $\mathcal{E}$. Obviously this is not case, but since $\mathcal{E}$ will be fixed throughout this notation simplifies the exposition.
\end{remark}

Although most of this note will focus on the behavioral implications of common belief in rationality, it will be useful to define the appropriate version of \emph{Bayes-Nash equilibrium} for a Bayesian game as introduced by \cite{harsanyi67}.\footnote{The epistemic foundation of Bayes-Nash equilibrium is still an open question. \citet[Theorem 15]{DS15} provide demanding sufficient conditions (beyond common belief in rationality) if there are two players.} For a given Bayesian game $\mathcal{B}$, define a \emph{strategy} of player $i$ by $s_i:\Theta_i \times Y_i \rightarrow A_i$.\footnote{I consider only pure strategies here in this note for simplicity and ease of notation. The analysis extends to mixed strategies.} Then a strategy profile $s=(s_1, s_2)$ constitutes a Bayes-Nash equilibrium if for every player $i$ and every $(\theta_i, y_i) \in \Theta_i \times Y_i$
\begin{align}\tag{BNE}
	s_i(\theta_i, y_i) \in BR_i(s_{-i} \circ \pi_i(\theta_i, y_i); \theta_i), \label{eq:BNE}
\end{align}
where $s_{-i} \circ \pi_i(\theta_i, y_i)[\theta_0, \theta_{-i}, a_{-i}] := \sum_{y_{-i} \in Y_{-i}\,:\, s_{-i}(\theta_{-i},y_{-i})=a_{-i}}\pi_i(\theta_i, y_i)[\theta_0, \theta_{-i}, y_{-i}]$. Let $BNE^\mathcal{Y}=(BNE_i^\mathcal{Y})_{i \in I}$ denote the set of Bayes-Nash equilibrium strategy profiles for information structure $\mathcal{Y}$.\footnote{\autoref{remark:infoonly_ICR} applies here as well.}


\subsection{Informational Robustness}
\label{subsection:3_inforobust_outside}

\cite{BM13, BM16, BM17} study the question of informational robust predictions in Bayesian games if the analyst (or any outside observer) does not know the information structure of the Bayesian game. Their starting point is Bayes-Nash equilibrium, whereas I take common belief in rationality as the starting point. Thus, the outside observer wants to characterize the set of all actions that are interim-correlated rationalizability for \emph{any} information structure. The following proposition formally relates this information robustness question to belief-free rationalizability.

\begin{proposition}\label{proposition:union_ICR_BFR}
		Let $\mathcal{E}$ be an economic environment. For every player $i$ and every payoff type $\theta_i$, $a_i \in BFR^\infty_i(\theta_i)$ if and only if there exists an information structure $\mathcal{Y}$ and a signal $y_i$ such that $a_i \in ICR_i^{\infty, \mathcal{Y}}(\theta_i, y_i)$. That is,
	\begin{align*}
		BFR^\infty_i(\theta_i) = \bigcup_{\mathcal{Y}} \bigcup_{y_i \in Y_i} ICR_i^{\infty, \mathcal{Y}}(\theta_i, y_i).
	\end{align*}
\end{proposition}

\begin{proof}
	Follows as a direct application of \autoref{proposition:union_ICR} with informational restrictions $\Delta_{i,\theta_i} = \Delta(\Theta_0 \times \Theta_{-i})$.
\end{proof}

\cite{BD87} establish an informational robustness interpretation similar to above for (complete information) rationalizability starting from Nash equilibrium. \citet[Propositions 4.2 and 4.3]{BS03} extend this robustness interpretation to Bayes-Nash equilibrium allowing for belief restrictions,\footnote{See \autoref{corollary:union_BNE} below.} whereas \citet[Section 4.5]{BM17} mention a similar interpretation for belief-free rationalizability, in particular. The following corollary establishes the formal result as conjectured in \cite{BM17}---it follows directly as a corollary from \autoref{proposition:union_ICR_BFR} and \citet[Remark 2]{DFM07}:

\begin{corollary}\label{corollary:union_BNE_BFR}
	Let $\mathcal{E}$ be an economic environment. For every player $i$ and every payoff type $\theta_i$, $a_i \in BFR^\infty_i(\theta_i)$ if and only if there exists an information structure $\mathcal{Y}$ and a Bayes-Nash equilibrium strategy profile $s$ 
	such that for every player $i$ and every payoff type $\theta_i$ there exists \emph{some} signal $y_i^* \in Y_i$ such that $a_i = s_i(\theta_i, y_i^*)$.
 That is,
	\begin{align*}
		BFR^\infty_i(\theta_i) = \bigcup_{\mathcal{Y}} \bigcup_{s_i \in BNE_i^\mathcal{Y}} \bigcup_{y_i \in Y_i} s(\theta_i, y_i).
	\end{align*}
\end{corollary}


\subsection{Localized Informational Robustness}
\label{subsection:3_inforobust_delta}
The previous section captured in a sense the robustest prediction to informational assumptions: the analyst does not want to make any assumptions about information (and higher-order beliefs about this assumption). Therefore, the previous results can be seen as \emph{global} informational robustness. However, in some instances this might be a too demanding robustness questions; maybe the analyst is willing to make some informational assumptions. For example, that every player assigns positive probability to all of the opponent's payoff types and this itself is commonly believed among the players. In such a case taking the union across \emph{all} information structures (and all signal realizations) is too permissive: the analyst wants to consider all information structures which are consistent with the baseline informational assumptions.

Following \cite{battigalli99,battigalli03} and \cite{BS03}, I formalize the baseline informational assumptions via belief restrictions for each player.\footnote{\cite{battigalli99,battigalli03} and \cite{BS03} consider more general belief restrictions; potentially also restricting beliefs about opponent's actions. To highlight the difference, I will refer to the belief restrictions in this informational setting as \emph{informational restrictions}. \cite{AKS13} and \cite{OP17} study robust mechanism design with belief restrictions via informational restrictions as considered here.} Informally, the analyst is willing to make the assumption that player's beliefs about the payoff state lie within a pre-specified set of beliefs. Formally, an \emph{informational restriction} is a pair $\Delta := (\Delta_1, \Delta_2) := \left(\left(\Delta_{1, \theta_1}\right)_{\theta_1 \in \Theta_1},\left(\Delta_{2, \theta_2}\right)_{\theta_2 \in \Theta_2}\right)$, where for each player $i$ and each payoff type $\theta_i$, $\Delta_{i, \theta_i}\subseteq \Delta(\Theta_0 \times \Theta_{-i})$ is a set of beliefs about the unknown parts of the payoff state, i.e. $\Theta_0 \times \Theta_{-i}$. The interpretation is that players (and payoff types) received some information about the overall payoff state, but the analyst is only willing to assume the resulting beliefs are within $\Delta_i$. If $\Delta_{i,\theta_i} = \Delta(\Theta_0 \times \Theta_{-i})$ for every $\theta_i$ and every player $i$ then there are no restrictions, which corresponds to the case analyzed in the previous part.

If the analyst is not concerned about robustness, she would---as before---model the strategic situation with a Bayesian game, but now she would make sure that the information structure is consistent with the hypothesized informational restrictions.

\begin{definition}
	For a given economic environment $\mathcal{E}$ and an informational restriction $\Delta$, an information structure $\mathcal{Y}$ is \emph{consistent} with $\Delta$ when  $\marg_{\Theta_0 \times \Theta_{-i}} \pi_i(\theta_i, y_i) \in \Delta_{i,\theta_i}$ for every information type $(\theta_i, y_i) \in \Theta_i \times Y_i$ and every player $i$. Let $\mathbb{Y}(\Delta)$ denote the collection of all information structures consistent with $\Delta$.
\end{definition}

The localized informational robustness question for a given informational restriction $\Delta$ is to characterize the set of all interim-correlated actions across all information structures consistent with $\Delta$. Unless $\Delta$ does not impose any restrictions it is obvious that the resulting set of actions will be a refinement of belief-free rationalizability. Indeed, as shown below the relevant robust solution concept is given by $\Delta$-rationalizability as introduced by \cite{battigalli99,battigalli03} and \cite{BS03}. As before the definition is given inductively: action $a_{i}$ is \emph{$\Delta$-rationalizable} for payoff type $\theta_i$ if it is iteratively a best-reply to a belief that ($i$) is inside the allowed beliefs $\Delta_{i,\theta_i}$ and ($ii$) assigns positive probability to actions of the other player that survived the previous round. Formally, action $a_{i}$ is $\Delta$-rationalizable for payoff type $\theta_i$ if $a_{i}\in \Delta R_{i}^{\infty}(\theta_i):=\bigcap_{n\geq 0}\Delta R_{i}^{n}(\theta_i)$, where $\Delta R_{i}^{0}(\theta_i):=A_{i}$ and inductively for any $n\in\mathds{N}$, 
\begin{align}\tag{$\Delta R^n$}\resizebox{0.8\textwidth}{!}{$
\Delta R_i^{n}(\theta_i):=\left\{a_{i}\in \Delta R_i^{n-1}(\theta_i)\left|
\begin{tabular}{r l}
\multicolumn{2}{l}{$\text{There exists }\mu_{i}\in\Delta(\Theta_0 \times \Theta_{-i} \times A_{-i})\text{ s.t.:}$}\\[1ex]
$(i)$&$\marg_{\Theta_0 \times \Theta_{-i}} \mu_{i} \in \Delta_{i,\theta_i} $,\\[1ex]
$(ii)$&$\supp \mu_{i}\subseteq \Theta_0 \times \graph\left( \Delta R_{-i}^{n-1} \right)$,\\[1ex]
$(iii)$&$a_{i}\in BR_{i}(\mu_{i}; \theta_i)$
\end{tabular}
\right.\right\}.$} \label{eq:DeltaR_iterate} 
\end{align}
As before a fixed-point definition of $\Delta$-rationalizability comes in handy: for every player $i$ and every payoff type $\theta_i$ consider a set of actions $F_i(\theta_i)$ with the following fixed-point property:
\begin{align}\tag{$\Delta R_{FP}$}\resizebox{0.875\textwidth}{!}{$
F_i(\theta_i):=\left\{a_{i}\in A_i\left|
\begin{tabular}{r l}
\multicolumn{2}{l}{$\text{There exists some  }\mu_{i}\in\Delta(\Theta_0 \times \Theta_{-i} \times A_{-i})\text{  such that:}$}\\[1ex]
$(i)$&$\marg_{\Theta_0 \times \Theta_{-i}} \mu_{i} \in \Delta_{i,\theta_i} $,\\[1ex]
$(ii)$&$\supp \mu_{i}\subseteq \Theta_0 \times \graph \left(F_{-i} \right)$,\\[1ex]
$(iii)$&$a_{i}\in BR_{i}(\mu_{i}; \theta_i)$
\end{tabular}
\right.\right\}.$} \label{eq:DeltaR_FP}
\end{align}
Then the pair $\left(\Delta R_i^\infty\right)_{i \in I}$ understood as correspondences is equal to the pair of correspondences $\left(F_i\right)_{i \in I}$ satisfying the fixed-point property and are largest by set inclusion.

With this definition in hand, the main result can be stated formally:
\begin{proposition}\label{proposition:union_ICR}
	Let $\mathcal{E}$ be an economic environment and fix belief restrictions $\Delta$. For every player $i$ and every payoff type $\theta_i$, $a_i \in \Delta R^\infty_i(\theta_i)$ if and only if there exists an information structure $\mathcal{Y}$ consistent with $\Delta$ such that for every player $i$ and every payoff type $\theta_i$ there exists \emph{some} signal $y_i \in Y_i$ such that $a_i \in ICR_i^{\infty, \mathcal{Y}}(\theta_i, y_i)$.
 That is,
	\begin{align*}
		\Delta R^\infty_i(\theta_i) = \bigcup_{\mathcal{Y} \in \mathbb{Y}(\Delta)} \bigcup_{y_i \in Y_i}  ICR_i^{\infty, \mathcal{Y}}(\theta_i, y_i).
	\end{align*}
\end{proposition}

\begin{proof}
	\emph{If.} For a given information structure $\mathcal{Y}$ such that $\marg_{\Theta_0 \times \Theta_{-i}} \pi_i(\theta_i,y_i) \in \Delta_{i, \theta_i}$   for every player $i$ and every information type $(\theta_i,y_i) \in \Theta_i \times Y_i$, consider a signal $y_i \in Y_i$ and an action $a_i \in A_i$ such that $a_i \in ICR_i^{\infty, \mathcal{Y}}(\theta_i, y_i)$. I show that  $a_i \in \Delta R_i^\infty(\theta_i)$ by induction, i.e.\! $a_i \in \Delta R_i^n(\theta_i)$ for every $n$. The statement is trivial for $n=0$. The induction hypothesis is that for $n\geq 0$, for any player $i$, and any information type $(\theta_i,y_i)$, $a_i \in ICR_i^{\infty, \mathcal{Y}}(\theta_i, y_i)$ implies $a_i \in \Delta R_i^n(\theta_i)$. Induction requires to verify this statement for $n+1$ too. Now, for $a_i \in ICR_i^{\infty, \mathcal{Y}}(\theta_i, y_i)$, \autoref{eq:ICR_FP} provides a belief $\mu_i \in \Delta(\Theta_0 \times \Theta_{-i} \times Y_{-i} \times A_{-i})$ such that
	\begin{align*}
		&(1) \, \supp \mu_{i}\subseteq \Theta_0 \times \left\{(\theta_{-i}, y_{-i}, a_{-i}) \in \Theta_{-i} \times Y_{-i}  \times A_{-i} \left| \, a_{-i} \in ICR^{\infty,\mathcal{Y}}_{-i}(\theta_{-i},y_{-i})\right.\right\}, \\
		&(2)\; a_i \in BR_{i}\left(\marg_{\Theta_0 \times \Theta_{-i} \times A_{-i}}\mu_{i}; \theta_i \right), \text{ and }\\
		&(3)\, \marg_{\Theta_0 \times \Theta_{-i} \times Y_{-i}}\mu_{i} = \pi_i(\theta_i,y_i).
	\end{align*}
	Let $\hat{\mu}_i:=\marg_{\Theta_0 \times \Theta_{-i} \times A_{-i}} \mu_i$, then (3) implies $\marg_{\Theta_0 \times \Theta_{-i}}\hat{\mu}_{i} \in \Delta_{i, \theta_i}$. Furthermore, for any $(\theta_0, \theta_{-i}, a_{-i})$ with $\hat{\mu}_i[\theta_0, \theta_{-i}, a_{-i}]> 0$, it holds that $\mu_i[\theta_0, \theta_{-i},y_{-i}, a_{-i}]>0$  for some  $y_{-i}$. Thus, by (1), $a_{-i} \in ICR_{-i}^{\infty, \mathcal{Y}}(\theta_{-i}, y_{-i})$. Now, by the induction hypothesis (applied for the other player) $a_{-i} \in \Delta R^{n}_{-i}(\theta_{-i})$. Finally, by (2), $a_i$ is also a maximizer for $\hat{\mu}_i$, so that $a_i \in \Delta R_i^{n+1}(\theta_i)$.
	
	\emph{Only If.} If $\Delta R^\infty_i(\theta_i)=\emptyset$ the statement is trivial. If not, then for every $a_i \in \Delta R^\infty_i(\theta_i)$, there is a justifying belief $\mu_i^{a_i, \theta_i} \in \Delta(\Theta_0 \times \Theta_{-i} \times A_{-i})$ satisfying ($i$)--($iii$) from \autoref{eq:DeltaR_FP}. Consider the following information structure: for every player $i$ set $Y_i = A_i$ and assign $\pi_i(\theta_i, a_i) = \mu_i^{a_i, \theta_i}$ for every information type $(\theta_i, a_i)$.
	For every payoff type $\theta_i$, ($i$) implies $\marg_{\Theta_0 \times \Theta_{-i}} \pi_i(\theta_i, y_i) \in \Delta_{i,\theta_i}$ for every $y_i \in Y_i$.
	Now, fix $a_i \in \Delta R^\infty_i(\theta_i)$ and I will prove by induction that $a_i \in ICR^{\infty,\mathcal{Y}}(\theta_i,a_i)$. The statement is trivial for $n=0$. The induction hypothesis is that for $n\geq 0$, for any player $i$, and any payoff type $\theta_i$, $a_i \in \Delta R_i^\infty(\theta_i)$  implies $a_i \in ICR_i^{n, \mathcal{Y}}(\theta_i, a_i)$. Induction requires to verify this statement for $n+1$ too. For this, again, consider arbitrary $a_i \in \Delta R^\infty_i(\theta_i)$ and define $\mu_i \in \Delta(\Theta_0 \times \Theta_{-i} \times Y_{-i} \times A_{-i})$ by
	\begin{align*}
		\mu_i[\theta_0, \theta_{-i}, y_{-i}, a_{-i}] = \mu_i^{a_i, \theta_i}[\theta_0, \theta_{-i}, a_{-i}]
	\end{align*}
	if $y_{-i} = a_{-i}$ and zero otherwise. By construction, ($ii$) and ($iii$) of \autoref{eq:ICR_iterate} are satisfied. For ($i$), consider $(\theta_0, \theta_{-i}, y_{-i}, a_{-i})$ such that $\mu_i[\theta_0, \theta_{-i}, y_{-i}, a_{-i}]>0$. Then, by construction, $y_{-i} = a_{-i}$ and $\mu_i^{a_i, \theta_i}[\theta_0, \theta_{-i}, a_{-i}]>0$. Thus, $(ii)$ of \autoref{eq:DeltaR_FP} implies $a_{-i} \in \Delta R_{-i}^\infty(\theta_{-i})$. Now, the induction hypothesis (applied for the other player) gives $a_{-i} \in  ICR_{-i}^{n,\mathcal{Y}}(\theta_{-i},a_{-i})$. Therefore, ($i$) of \autoref{eq:DeltaR_FP} holds too and I conclude that $a_i \in ICR_i^{n+1,\mathcal{Y}}(\theta_i,a_i)$.
\end{proof}

As before, exploiting the relationship of ICR with Bayes-Nash equilibrium gives local informational robustness with equilibrium as baseline assumption as a corollary.
\begin{corollary}[\citealp{battigalli03, BS03}]\label{corollary:union_BNE}
	Let $\mathcal{E}$ be an economic environment and fix belief restrictions $\Delta$. For every player $i$ and every payoff type $\theta_i$, $a_i \in \Delta R^\infty_i(\theta_i)$ if and only if there exists an information structure $\mathcal{Y}$ consistent with $\Delta$ and a Bayes-Nash equilibrium strategy profile $s$
	such that for every player $i$ and every payoff type $\theta_i$ there exists \emph{some} signal $y_i \in Y_i$ such that $a_i = s_i(\theta_i, y_i)$.
 That is,
	\begin{align*}
		\Delta R^\infty_i(\theta_i) = \bigcup_{\mathcal{Y}\in \mathbb{Y}(\Delta)} \bigcup_{s_i \in BNE_i^\mathcal{Y}} \bigcup_{y_i \in Y_i}  s_i(\theta_i, y_i).
	\end{align*}
\end{corollary}

The main result about informational robustness of \citet[Proposition 8]{BM17} can be seen as a special case of \autoref{corollary:union_BNE}. In their model, there are no private payoff types. In my setting this corresponds to assuming that the utility functions $u_i$ depend only on $\theta_0$ (and on the action profile). With this assumption, $\theta_i$ is directly payoff irrelevant, but could potentially serve as a payoff-relevant signal (i.e.\! about $\theta_0$). Indeed, \citet{BM17} start with a baseline information structure with beliefs given by $\varphi_i: \Theta_i \rightarrow \Delta(\Theta_0 \times \Theta_{-i})$. Defining corresponding informational restrictions as $\Delta_{i,\theta_i} = \left\{ \mu_i \in \Delta(\Theta_0 \times \Theta_{-i}): \supp \mu_i \subseteq \supp \varphi_i(\theta_i)\right\}$ then yields \citeauthor{BM17}'s solution concept as $\Delta$-rationalizability using the definition in this note. \citeauthor{BM17} are interested in when players might have more information than the baseline information embodied in $\varphi$s. More information might render some states irrelevant by Bayesian updating, but it cannot make states probably when they were impossible according to the baseline information. With this consideration in mind, an analyst is only willing to make the informational assumptions given by the support of $\varphi$. Whereas \autoref{corollary:union_BNE} reduces to Proposition 8 of \citet{BM17}, \autoref{proposition:union_ICR} shows that their informational robustness question has the same answer when the starting point of the analysis is common belief in rationality instead of Bayes-Nash equilibrium.

\bibliographystyle{ecta} 
\bibliography{biblio}


\end{document}